# Infrared Spectroscopy of Wafer-Scale Graphene

Hugen Yan, Fengnian Xia[*], Wenjuan Zhu, Marcus Freitag, Christos Dimitrakopoulos,

Ageeth A. Bol, George Tulevski, and Phaedon Avouris[*]

*IBM Thomas J. Watson Research Center, Yorktown Heights, NY 10598*

**ABSTRACT**. We report on spectroscopy results from the mid- to far-infrared on wafer-scale graphene, grown either epitaxially on silicon carbide, or by chemical vapor deposition. The free carrier absorption (Drude peak) is simultaneously obtained with the universal optical conductivity (due to interband transitions), and the wavelength at which Pauli blocking occurs due to band filling. From these the graphene layer number, doping level, sheet resistivity, carrier mobility, and scattering rate can be inferred. The mid-IR absorption of epitaxial two-layer graphene shows a less pronounced peak at $0.37\pm0.02$ eV compared to that in exfoliated bilayer graphene. In heavily chemically-doped single layer graphene, a record high transmission reduction due to free carriers approaching 40% at 250 $\mu$m (40 cm$^{-1}$) is measured in this atomically thin material, supporting the great potential of graphene in far-infrared and terahertz optoelectronics.

*To whom correspondence should be addressed.
Email: fxia@us.ibm.com (F.X.); avouris@us.ibm.com (P.A.).

Graphene is a promising material for photonic and optoelectronic applications, due to its strong interaction with light in a broad wavelength range.[1-13] Graphene exhibits universal optical conductivity $e^2/4\hbar$ resulting from interband transitions, where $e$ is the electron charge and $\hbar$ is the Plank constant, leading to 2.3% absorption for vertical incidence photons in free standing graphene from visible to infrared.[2, 14] Moreover, due to Pauli blocking, the interband absorption of photons with energy below $2|E_F|$, where $E_F$ is the energy difference in Fermi level and Dirac point, is suppressed.[4] In the far-IR and terahertz regions, however, intraband transitions or free carriers dominate. The frequency dependence of free carrier response in graphene can be described by the Drude model using the dynamical conductivity $\sigma(\omega) = \dfrac{iD}{\pi(\omega + i\Gamma)}$, where $D$ is the Drude weight and $\Gamma$ is the carrier scattering rate.[10, 11] The Drude weight $D$, expressed as $\dfrac{v_f e^2}{\hbar}\sqrt{\pi|n|}$ for massless Dirac Fermions in single layer graphene, where $v_f$ is the Fermi velocity, has different dependence on the carrier concentration $n$ from that of the conventional massive Fermions.

As pristine graphene is gapless and a tunable, moderate band gap can be engineered using symmetry breaking schemes, it exhibits particularly strong potential in far-IR and terahertz optoelectronics.[15] Characterization of wafer-scale graphene in the IR and terahertz range is a crucial step in the development of graphene optoelectronic devices. Previously, broadband optical absorption measurements of few and multilayer epitaxial graphene down to the terahertz range provided useful information of layer number as well as doping levels.[16, 17] Large area graphene samples produced by chemical vapor-phase deposition (CVD) were also studied.[18-20] The free carrier response of back-gated CVD graphene devices in the far-IR was carefully examined by Horng *et al*.[18] for carrier densities below $7\times10^{12}$ cm$^{-2}$. A reduction of Drude weight was observed.

**RESULTS AND DISCUSSION**

In this letter, we report our comprehensive studies of few-layer wafer-scale epitaxial and CVD graphene by infrared spectroscopy. The epitaxial graphene samples were grown through silicon sublimation of 2-inch silicon carbide (SiC) wafers on silicon-face at a temperature of 1550$^{o}$C under argon atmosphere with a chamber pressure of 3.5 mTorr.[21] The thickness of the graphene layers was controlled by the graphenization time, which was 2 minutes for sample 1 and 10 minutes for sample 2. Figs. 1 (a) and (b) show the atomic force microscopic (AFM) images of the samples 1 and 2, respectively. Sample 1 mainly consists of single layer graphene, with some bilayer and blank areas. Sample 2 is covered mostly by bilayer graphene with small trilayer areas. Before the infrared spectroscopy measurement, the carbon-face graphene on the back of the SiC substrate was removed through a brief exposure to oxygen plasma. Single layer CVD graphene was grown on copper foil and subsequently transferred to a quartz substrate.[22] Fig. 1 (c) is the microscope image of the transferred graphene on quartz substrate. The size of the graphene is about 1.5 cm by 1.5 cm. The Raman spectrum of the CVD graphene is shown in Fig. 1(d), and the small D mode to G mode intensity ratio indicates the high quality of this CVD graphene.

The spectra of reduction in transmission for epitaxial graphene samples 1 and 2 are shown in Fig. 2 with the inset focusing on the mid-IR range. We were not able to measure the absorption spectra in the restrahlen band (600cm$^{-1}$ to 1800cm$^{-1}$) of SiC due to the near zero transmission. Some small but sharp features at near 2900 cm$^{-1}$ come from C-H vibration bands due to water adsorption on graphene.[23] They also show up in CVD graphene absorption spectra (Fig. 3 inset). For both samples, the absorption is rather flat in a broad range in the mid-IR. This is a consequence of interband transitions of Dirac Fermions, which result in a frequency independent optical conductivity. The value of the absorption is proportional to the average layer number of the graphene and depends on the refractive index of the substrate as well. By applying equation (3) in the METHODS section with $n_s$=2.55 and fitting the relatively flat part of the data (the red curves in the mid-IR range of Fig. 2 are fitting curves), we obtain the layer number $N$ = 1.5±0.2 for sample 1 and $N$ = 2.0±0.2 for sample 2. This is in good agreement with AFM images shown in Figs. 1 (a) and (b), although the AFM images only provide local information. Due to Pauli blocking, both samples 1 and 2 show reduced absorption in the lower wave number region (below 2700cm$^{-1}$) as shown in Fig. 2. In these two cases, we are unable to estimate the Fermi level from the Pauli blocking, because of the restrahlen band of the SiC substrate. However, with the carrier density obtained from the far-IR measurements as mentioned later, we could fit each absorption curve in the mid-IR range with a Gaussian broadened step function (red curves in Fig. 2).

A notable deviation from the fitting for sample 2 is the broad peak at 0.37 ±0.02eV (2850±150cm$^{-1}$), indicated by the green arrow in the inset. Similar absorption peaks in the mid-IR, usually much stronger and sharper, were previously observed in exfoliated bilayer and multilayer graphene.[9, 13] They are associated with the different band structures of the multilayer graphene flakes. Bernal-stacked bilayer graphene shows an absorption peak at around 0.37eV, and the peak absorption can reach about two times of the universal value, depending on the doping level.[9, 24] We attribute the broad peak in Fig. 2 to the existence of the Bernal-stacked bilayer graphene on the epitaxial wafer. However, the peak absorption is only about 10% larger compared with the universal value. Multilayer epitaxial graphene on Si-face is generally believed to have Bernal (AB) stacking,[25] which is quite different from C-face graphene. Scanning tunneling microscopic studies show deviations from Bernal stacking in some areas.[26, 27] Our results suggest that only a small fraction of the bilayer epitaxial graphene from the Si-face has perfect Bernal (AB) stacking. This is consistent with Raman measurements which rarely show Bernal stacking signature in the 2D mode for this sample and other similar bilayer samples (see supporting information).[28] The Raman 2D mode is a widely used fingerprint to unambiguously identify Bernal-stacked bilayer graphene prepared through exfoliation.[29] Recent transport measurements indicate the existence of a band gap in biased bilayer graphene on SiC much smaller than that in exfoliated graphene.[30] Our infrared spectroscopy results are consistent with these transport observations. The exact nature of the stacking and interlayer coupling for epitaxial multilayer graphene needs further investigation. In single layer epitaxial graphene extinction spectrum, there exists a weak and broad peak at around 4500 cm$^{-1}$, which also varies from sample to sample. The peak-to-valley amplitude is only around 0.1%. The possible causes can be the non-uniformity in the samples and uncertainty in the measurements.

In the far-IR region, the absorption increases rapidly with decreasing frequency in both samples due to free carrier absorption as shown in Fig. 2. The Drude conductivity is

$$\sigma(\omega) = \frac{iD}{\pi(\omega + i\Gamma)} \qquad (1)$$

We fit the data using equations (1) and (2) (see METHODS), with $D$ and $\Gamma$ as fitting parameters. Note that the approximated expression (eqn. (3)) was not used since the Drude absorption is more pronounced and both the real and imaginary parts of the Drude conductivity have to be taken into account. The red curves in the far-IR are the fitting results with $D = 6.3 \times 10^3$ $e^2/h$ cm$^{-1}$, $\Gamma$=269 cm$^{-1}$ for sample 1, and $D = 1.2 \times 10^4$ $e^2/h$ cm$^{-1}$, $\Gamma$ = 239 cm$^{-1}$ for sample 2. It should be noted that the Drude weight depends on the thickness of the multilayer graphene. Since our graphene samples are 1-2 layers thick and only a small fraction is Bernal-stacked, it is suitable to assume that the carriers are distributed equally in each independent layer.[31] As a consequence, $D = N \frac{v_f e^2}{\hbar} \sqrt{\pi |n| / N} = \frac{v_f e^2}{\hbar} \sqrt{\pi N |n|}$ , where $N$ is the layer numbers. With this, carrier densities of $0.6 \times 10^{12}$ cm$^{-2}$ for sample 1 and $1.6 \times 10^{12}$ cm$^{-2}$ for sample 2 can be derived. The carrier type cannot be determined from the infrared measurement. Furthermore, by extrapolating the optical conductivity to zero frequency, the DC conductivity can be obtained. The DC mobility can then be calculated from the carrier density and the conductivity. The mobilities thus obtained are 2940 cm$^2$/Vs for sample 1 and 2350 cm$^2$/Vs for sample 2. We also performed Hall measurements in a Van der Pauw configuration to check the carrier density and mobility. We cut samples 1 and 2 into 1 cm by 1 cm squares. To avoid extra doping commonly introduced in device fabrication processes, those two samples were directly contacted without processing. Both samples were found to be electron doped and the measured carrier densities are $0.8 \times 10^{12} \text{cm}^{-2}$ and $1.7 \times 10^{12} \text{cm}^{-2}$ for samples 1 and 2 respectively, in good agreement with the values obtained from the far-IR measurements. The measured Hall-mobilities are 783 $\text{cm}^2/\text{Vs}$ for sample 1 and 1360$\text{cm}^2/\text{Vs}$ for sample 2. The extrapolated mobilities from the far-IR measurements are larger than those obtained using Hall measurements, especially for sample 1. This discrepancy is largely due to the fact that any discontinuity of the 1cm by 1cm graphene film can strongly affect the DC transport mobility but has smaller effect on the optical conductivity in the terahertz range. Thinner samples are likely to have more discontinuities, which can explain the larger discrepancy observed for sample 1. Besides, the mobilities from the van der Pauw measurements are averaged over the whole 1cm by 1cm wafer, while the infrared measurements are for 2 mm size areas. Nevertheless, the derived mobility from the Drude conductivity in the far-IR is very close to the measured mobility through transport on micron-sized samples used in most of the transport studies.

Besides the epitaxial graphene, large area CVD graphene was also investigated. Fig. 3 shows the absorption spectrum of CVD graphene on quartz. The spectra from 250 to 270 $\text{cm}^{-1}$ and 320 to 2100 $\text{cm}^{-1}$ are not accessible due to the strong absorption of the quartz substrate in these wavelength ranges. With decreasing photon frequency in the mid-IR, the absorption decreases from 1.9% to 0.4% due to Pauli blocking; in the far-IR, the absorption increases rapidly to 25% at 40$\text{cm}^{-1}$ (1.2 THz in frequency, or 250 μm in wavelength) due to free carrier absorption. The inset of Fig. 3 is an enlarged view of the mid-IR spectrum. The blue line is the universal interband absorption of a single layer graphene on quartz substrate (index of refraction $n_s = 1.44$ for quartz in mid-IR). The red curve is a fitted absorption spectrum using a phenomenological Gaussian broadened step function which includes doping inhomogeneity, temperature effect and other broadening mechanisms. The full-width-half-maximum (FWHM) of the Gaussian function is 2001 cm$^{-1}$ and centered at 5501.6 cm$^{-1}$, corresponding to |E$_F$| of 341 meV (or a carrier density $n$ of $7.1 \times 10^{12}$ cm$^{-2}$). The Fermi level was further confirmed by Raman spectroscopy. A typical Raman spectrum in Fig. 1 (d) shows that both the G and 2D modes are stiffened if compared with those in intrinsic graphene. The G mode frequency can be used to estimate the Fermi level:[32, 33]

$$\omega_G - 1580 cm^{-1} = (42 cm^{-1} / eV) \times \left| E_F \right| \qquad (4)$$

In addition, the stiffening of the 2D mode is an indication of hole doping.[34] We mapped an 80 μm by 60 μm area and obtained an average Fermi level of -321 meV (see supporting information), this is in good agreement with the Fermi level estimated from the location of the Pauli blocking. Besides, Raman mapping also indicates that the doping is inhomogeneous across the sample. This has considerable contribution to the broadening of the absorption step at 2 |E$_F$| (see supporting information).

In the far-IR region, the Drude model describes the dependence of the absorption on frequency accurately, as shown in Fig. 3. Note that the index of refraction of quartz in the far-IR region is 2.14,[35] which is different from that in mid-IR. The red curve in Fig. 3 is the fitting curve using the Drude model with fitting parameters $D$=1.24 ×10$^4$ $e^2/h$ cm$^{-1}$ and Γ=102 cm$^{-1}$. By extrapolating the Drude model to zero frequency, we obtain the DC sheet resistance ρ$_s$ of 670 Ω/□. Furthermore, in conjunction with the carrier density obtained from the location of Pauli blocking (or Raman scattering), a hole mobility of 1300

cm$^2$/Vs is obtained. Both the sheet resistance and mobility are among typical values obtained through transport measurements for similar CVD graphene samples.[36] The Drude weight $D$, can also be calculated using $D = \frac{v_f e^2}{\hbar} \sqrt{\pi |n|}$ since the carrier density can be derived from the location of Pauli blocking. The calculated D is $1.7 \times 10^4$ $e^2/h$ cm$^{-1}$ and the directly measured $D$ through the far-IR absorption is about 28% smaller. This is consistent with the recent findings by Horng *et al.*, although the carrier density is higher in our case.[18] The mechanisms for Drude weight reduction are not clear at this stage. One of the theoretical studies even predicts Drude weight enhancement if electron-electron interactions are taken into account.[37] It should also be noted that the reduction of Drude weight varies from sample to sample and some samples have no reduction at all.

To study the infrared response of graphene under extremely high carrier density, the CVD graphene was soaked into the solution of a single electron oxidant OA ($(C_2H_5)_3OSbCl_6$) for half an hour to introduce extra hole doping,[38, 39] Fig. 4 (b) shows the G mode Raman spectra before and after the chemical doping. An up-shift is observed, from which a hole density of $1.9 \times 10^{13}$ cm$^{-2}$ is inferred, almost 3 times greater than that of the as-prepared CVD graphene. Corresponding changes in the infrared absorption spectra are presented in Fig. 4a. First, as depicted in the inset, the absorption in the entire mid-IR region is suppressed due to Pauli blocking and the $2|E_F|$ is beyond our measurement range. This is consistent with the Raman scattering measurements, from which $2|E_F|$ of 1.12 eV (8929 cm$^{-1}$) is inferred. Second, the far-IR absorption increases to 37% at 40 cm$^{-1}$ and Drude model (red curve) can still be used to describe the absorption in the far-IR perfectly, with fitting parameters $D = 1.8 \times 10^4 \, e^2/h \, \text{cm}^{-1}$ and $\Gamma = 91 \, \text{cm}^{-1}$. In this case, the Drude weight obtained from far-IR absorption measurements is about 35% smaller if compared to the theoretical value ($D = 2.8 \times 10^4 \, e^2/h \, \text{cm}^{-1}$).

Finally, a few comments should be made to the carrier scattering rate $\Gamma$. The two epitaxial samples have similar scattering rate of ~270 cm$^{-1}$, which corresponds to a scattering time of 20 femto-seconds. The CVD graphene has smaller scattering rate of ~100 cm$^{-1}$ (scattering time of ~50 femto-seconds) and it doesn't increase with the presence of chemical absorbates from the chemical doping. The difference for the scattering rate between epitaxial and CVD graphene can be due to the different sample quality, the surface morphology, and the doping concentration. The scattering times measured here are consistent with transport measurements.[40]

**CONCLUSIONS**

To conclude, we demonstrate that infrared spectroscopy is an excellent technique to characterize wafer-scale graphene in a non-invasive manner. We show that through the analysis of the absorption spectra in mid- and far-IR ranges simultaneously, layer number, doping, sheet resistivity, carrier mobility, and scattering rate can be inferred. For chemically doped CVD graphene, the reduction of transmission in far-IR in this atomically thin film approaches 40%, demonstrating the great potential of this novel two dimensional material in far-IR and terahertz applications, such as terahertz detectors and imagers.

**METHODS**

Nicolet 8700 FT-IR spectrometer was used to perform the absorption measurement in a transmission geometry. Two detectors, liquid nitrogen cooled mercury-cadmium-telluride (MCT) detector and liquid helium cooled bolometer, were used to cover mid-IR (650 cm$^{-1}$ to 7500 cm$^{-1}$) and far-IR (40 cm$^{-1}$ to 700 cm$^{-1}$) regions, in conjunction with potassium-bromide (KBr) and solid-substrate beam splitters, respectively. We measured the transmission and reference spectra ($T$ and $T_0$) through the substrate with graphene and the bare substrate. For epitaxial graphene, a blank silicon carbide wafer was used for reference purposes. The vertical incidence infrared beam size on the sample is about 2 millimeters in diameter. All measurements were done at room temperature in nitrogen environment. For graphene on a substrate with index of refraction $n_s$, the reduction in transmission 1-T/T$_0$ is related to the complex optical conductivity $\sigma(\omega)=\sigma'(\omega)+i\sigma''(\omega)$ in the following way:[16, 17]

$$1-T/T_0 = 1 - \frac{1}{\left|1+Z_0\sigma(\omega)/(1+n_s)\right|^2} \qquad (2)$$

where $Z_0$ is the vacuum impedance $\sqrt{\mu_0/\varepsilon_0}$ and $\omega$ is the frequency. When the reduction in transmission is small, it only depends on the real part of the conductivity $\sigma'$:

$$1-T/T_0 = \frac{2Z_0\sigma'(\omega)}{1+n_s} \qquad (3)$$

Therefore, through the spectroscopy measurement, we can derive the real part of the optical conductivity in a broad spectral range. We use the Drude conductivity for the far-IR region and the universal optical conductivity and Pauli blocking in the mid-IR to analyze the data.

*Acknowledgments*

The authors are grateful to B. Ek and J. Bucchignano for help in technical support, and Chun-yung Sung for helpful discussions. H. Y. acknowledges the assistance on the bolometer from Dr. Zhiqiang Li (Columbia University). Correspondence should be addressed to F.X. (fxia@us.ibm.com) or P.A. (avouris@us.ibm.com).

*Supporting Information Available*: Raman mapping of CVD graphene, statistical analysis of the doping inhomogeneity, and Raman spectra of epitaxial graphene are presented in the supporting information. This material is available free of charge *via* the Internet at http://pubs.acs.org.

Figure 1 Microscopic characterization of wafer-scale graphene

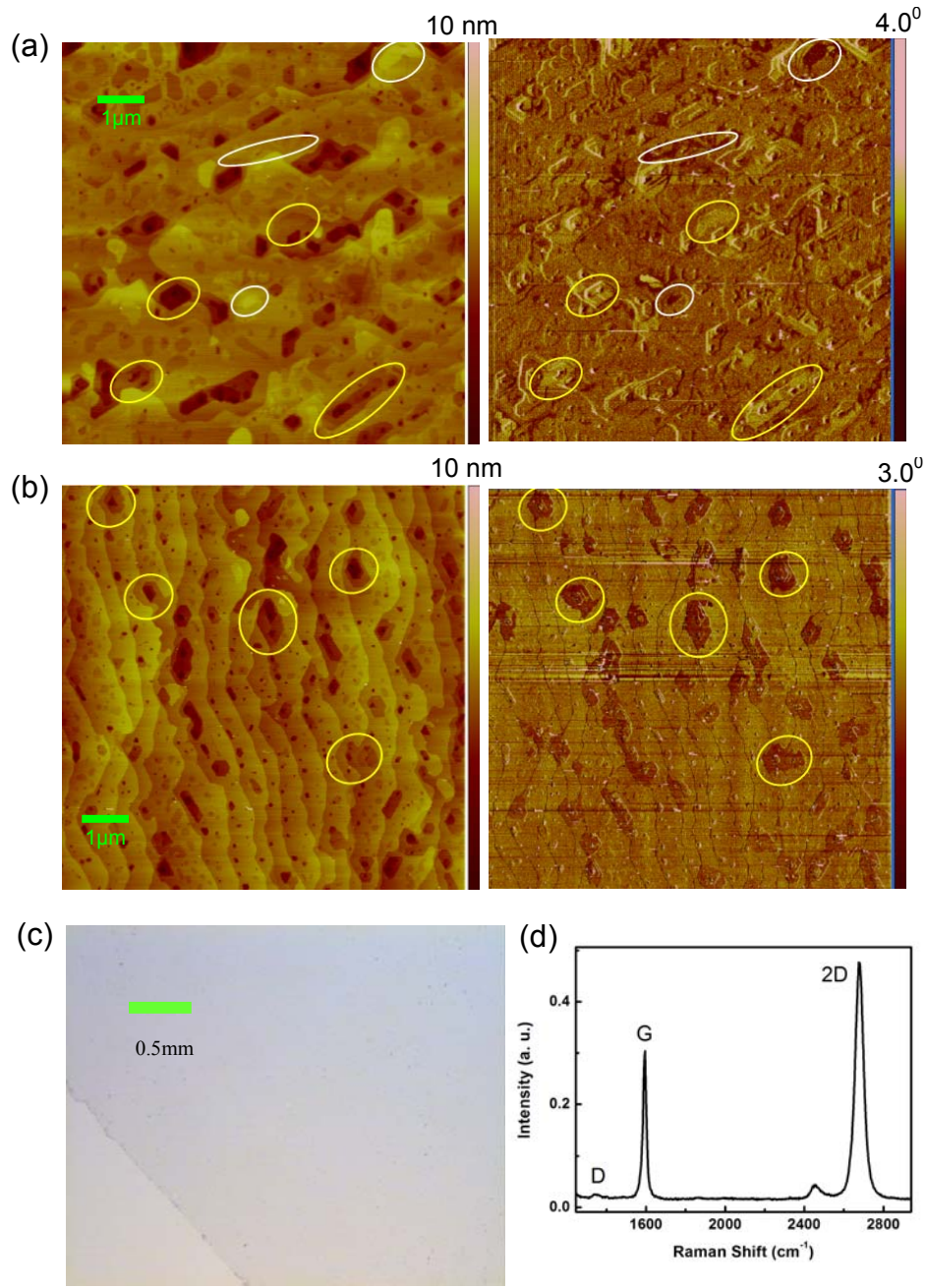

(a) and (b): Atomic force microscopic images in height (left) and phase mode (right) of

the epitaxial graphene samples 1 (a) and 2 (b). The median brown regions in the AFM phase image (on the right) of (a) indicates single layer graphene. Small area of dark brown regions, some marked by white ellipses, are buffer layer. The light brown regions, some of them marked with yellow ellipses, consist of two layers of graphene. The majority of the area in the phase image of (b) (light yellow color) is covered by two layers of graphene. Small area of trilayer graphene (dark brown region, some of them marked with ellipses) is also visible. (c): The microscopic image of the CVD graphene on quartz. (d): The Raman spectrum of the CVD graphene shown in (c). D, G and 2D modes are indicated.

Figure 2 Infrared absorption spectra of epitaxial graphene

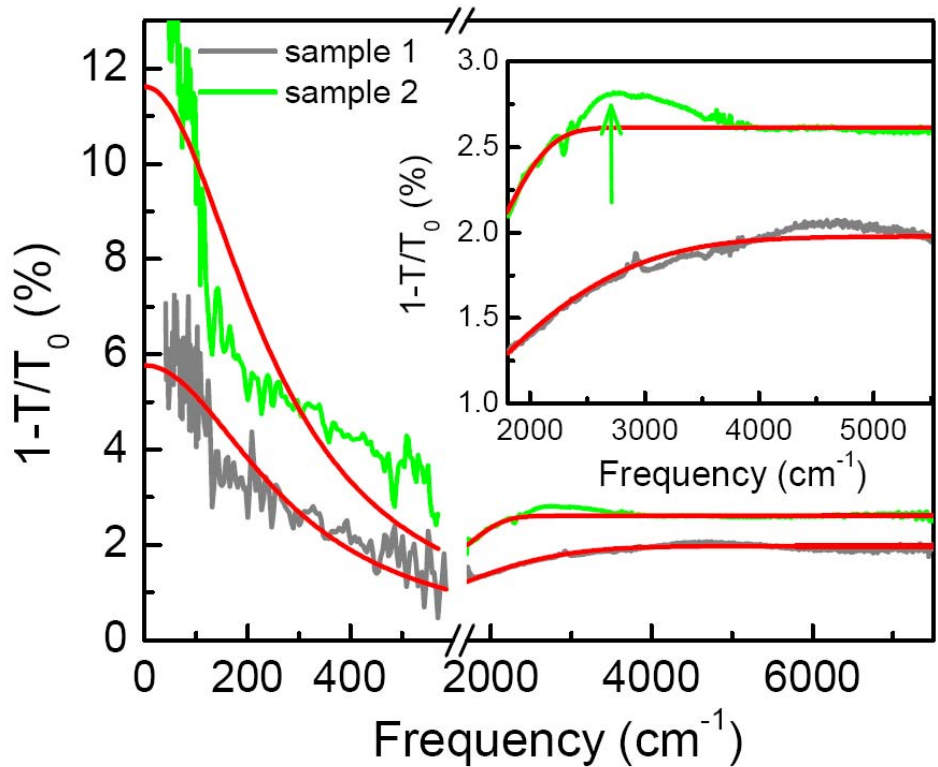

The red curves in far-IR region are calculated results with Drude model using parameters mentioned in the main text. Inset: the enlarged view of the mid-IR absorption curves. The red curves in mid-IR denote the values calculated using universal optical conductivity and Paul blocking with layer number $N = 1.5$ and 2 for sample 1 and 2, respectively. The green arrow in the inset points to the absorption peak from Bernal-stacked bilayer graphene covering the surface partially.

Figure 3 Infrared absorption spectra of CVD graphene on quartz

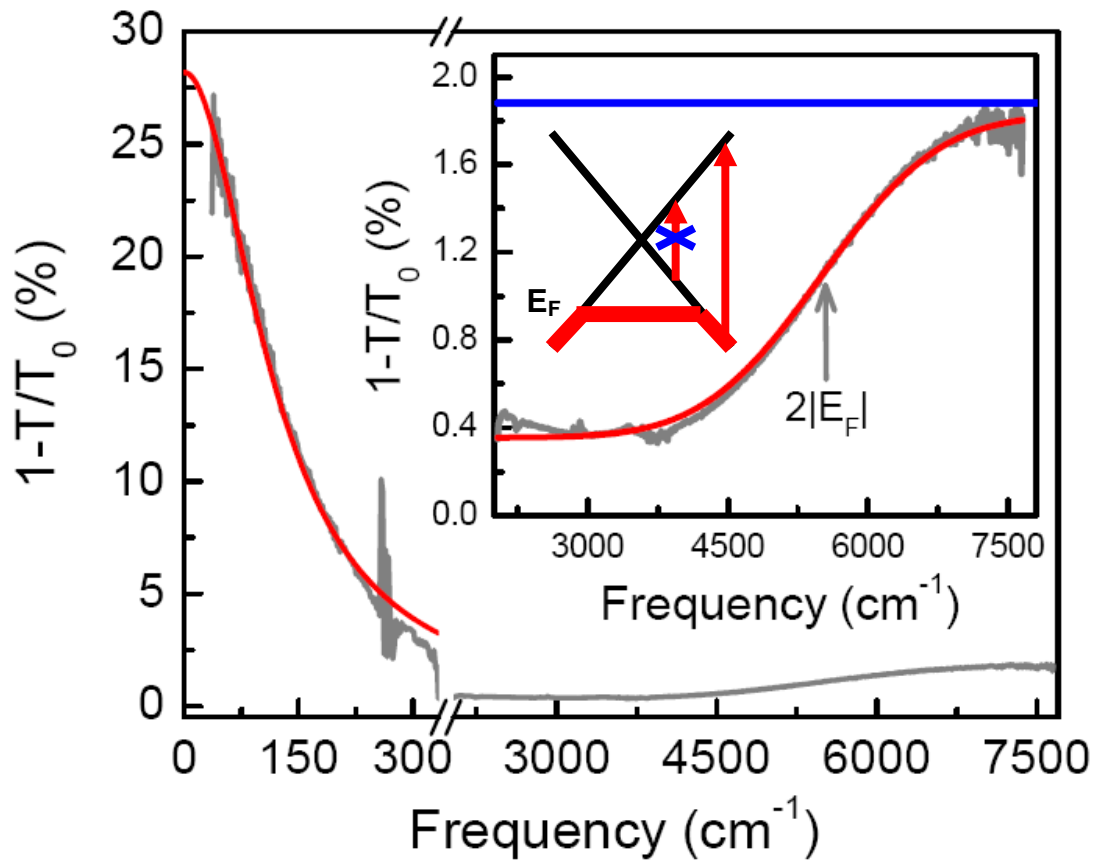

The red curve in far-IR is the fitting based on Drude model using parameters presented in the main text. Less noise in the far-IR spectrum than that in Fig. 2 is due to the higher light source power used. Inset: the enlarged view of the spectrum in mid-IR region with a sketch for the interband transitions. Red curve is the fitting based on Pauli blocking in mid-IR. The blue line is the universal interband absorption for single layer graphene on quartz. $2|E_F|$ is indicated by the grey arrow. The Dirac cone of graphene is also shown in the inset. Photons with energy smaller than $2|E_F|$ can not be absorbed by graphene due to Pauli blocking.

Figure 4 Impact of chemical doping on the infrared absorption spectrum of CVD graphene

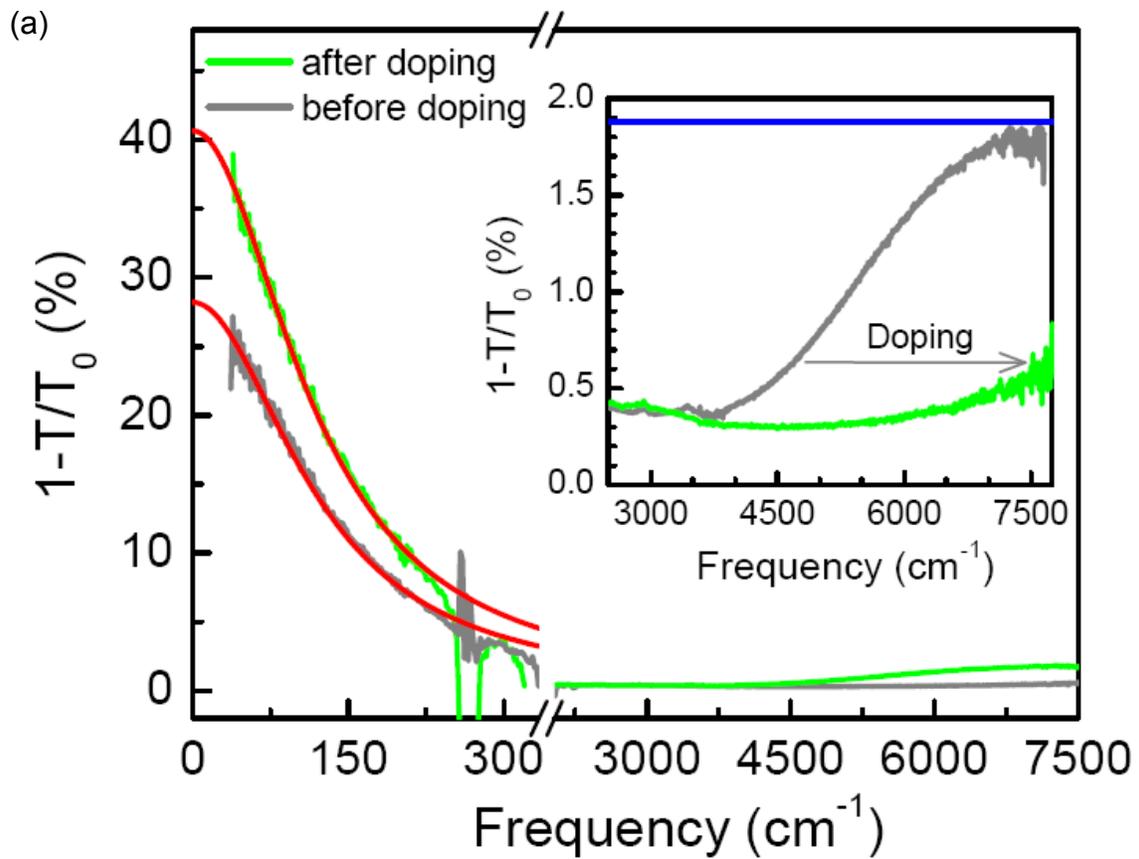

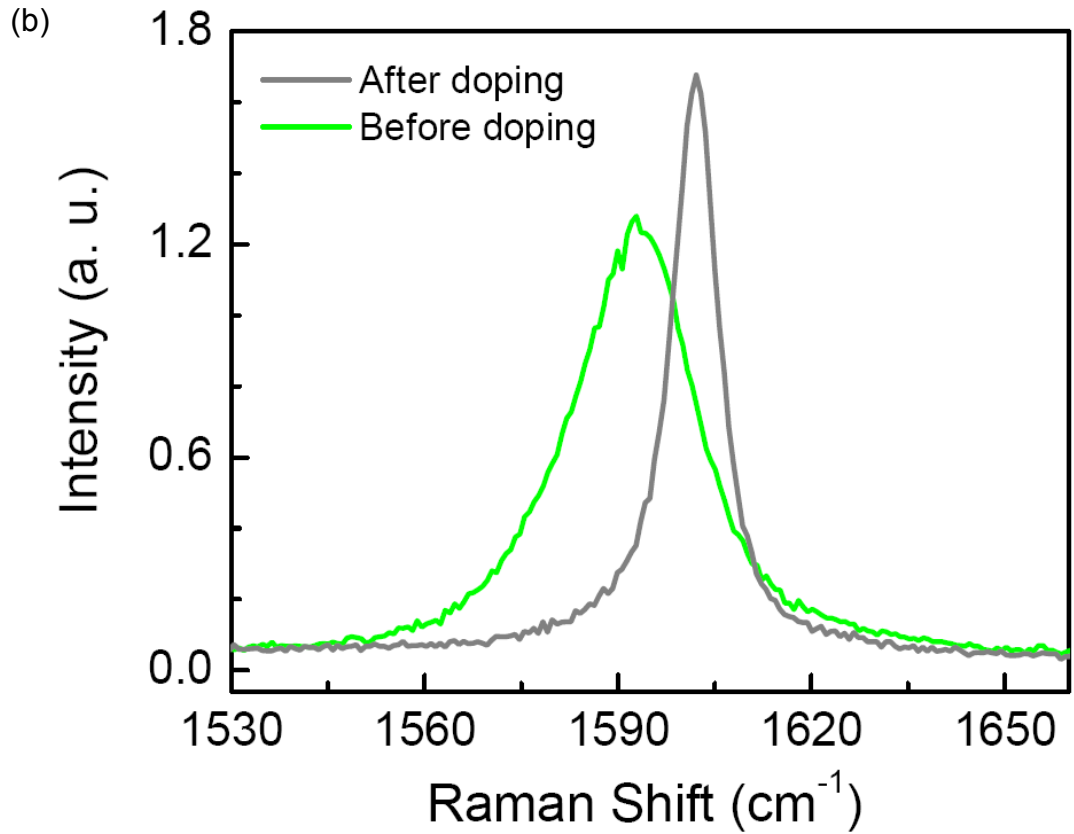

(b)

(a) Infrared absorption spectra for the CVD graphene before (green) and after (black) chemical doping. Red curves are fitting results using Drude model. Inset: the enlarged view in the mid-IR region. The blue line is the universal interband absorption for single layer graphene on quartz. (b) G mode Raman spectra for the CVD graphene before (green) and after (grey) chemical doping, from which the doping concentration can be estimated.